# Benchmarking of eight recurrent neural network variants for breath phase and adventitious sound detection on a self-developed open-access lung sound database—HF_Lung_V1

Fu-Shun Hsu[1,2,3], Shang-Ran Huang[3], Chien-Wen Huang[4], Chao-Jung Huang[3], Yuan-Ren Cheng[3,5,6], Chun-Chieh Chen[4], Jack Hsiao[7], Chung-Wei Chen[2], Li-Chin Chen[8], Yen-Chun Lai[3], Bi-Fang Hsu[3], Nian-Jhen Lin[3,9], Wan-Lin Tsai[3], Yi-Lin Wu[3], Tzu-Ling Tseng[3], Ching-Ting Tseng[3], Yi-Tsun Chen[3], Feipei Lai[1,†]

[1] Graduate Institute of Biomedical Electronics and Bioinformatics, National Taiwan University, Taipei 10617, Taiwan

[2] Department of Critical Care Medicine, Far Eastern Memorial Hospital, New Taipei 22060, Taiwan

[3] Heroic Faith Medical Science Co. Ltd., New Taipei 23553, Taiwan

[4] Avalanche Computing Inc., Taipei 10687, Taiwan

[5] Department of Life Science, College of Life Science, National Taiwan University, Taipei 10617, Taiwan

[6] Institute of Biomedical Sciences, Academia Sinica, Taipei 11529, Taiwan

[7] HCC Healthcare Group, New Taipei 22060, Taiwan

[8] Research Center for Information Technology Innovation, Academia Sinica, Taipei 11529, Taiwan

[9] Division of Pulmonary Medicine, Far Eastern Memorial Hospital, New Taipei 22060, Taiwan

Short Title: Automated lung sound analysis database

[†]Corresponding Author

Feipei Lai

Graduate Institute of Biomedical Electronics and Bioinformatics

National Taiwan University

No. 1, Sec. 4, Roosevelt Road

Taipei 10617, Taiwan

Tel: 886-2-3366-4961

Fax: 886-2-3366-3754

E-mail: flai@csie.ntu.edu.tw

## ABSTRACT

A reliable, remote, and continuous real-time respiratory sound monitor with automated respiratory sound analysis ability is urgently required in many clinical scenarios—such as in monitoring disease progression of coronavirus disease 2019—to replace conventional auscultation with a handheld stethoscope. However, a robust computerized respiratory sound analysis algorithm has not yet been validated in practical applications. In this study, we developed a lung sound database (HF_Lung_V1) comprising 9,765 audio files of lung sounds (duration of 15 s each), 34,095 inhalation labels, 18,349 exhalation labels, 13,883 continuous adventitious sound (CAS) labels (comprising 8,457 wheeze labels, 686 stridor labels, and 4,740 rhonchi labels), and 15,606 discontinuous adventitious sound labels (all crackles). We conducted benchmark tests for long short-term memory (LSTM), gated recurrent unit (GRU), bidirectional LSTM (BiLSTM), bidirectional GRU (BiGRU), convolutional neural network (CNN)-LSTM, CNN-GRU, CNN-BiLSTM, and CNN-BiGRU models for breath phase detection and adventitious sound detection. We also conducted a performance comparison between the LSTM-based and GRU-based models, between unidirectional and bidirectional models, and between models with and without a CNN. The results revealed that these models exhibited adequate performance in lung sound analysis. The GRU-based models outperformed, in terms of *F1* scores and areas under the receiver operating characteristic curves, the LSTM-based models in most of the defined tasks. Furthermore, all bidirectional models outperformed their unidirectional counterparts. Finally, the addition of a CNN improved the accuracy of lung sound analysis, especially in the CAS detection tasks.

## 1. Introduction

Respiration is vital for the normal functioning of the human body. Therefore, clinical physicians are frequently required to examine respiratory conditions. Respiratory auscultation (Bohadana et al., 2014; Goettel & Herrmann, 2019; Sarkar et al., 2015) using a stethoscope has long been a crucial first-line physical examination. The chestpiece of a stethoscope is usually placed on a patient's chest or back for lung sound auscultation or over the patient's tracheal region for tracheal sound auscultation. During auscultation, breath cycles can be inferred, which help clinical physicians evaluate the patient's respiratory rate. In addition, pulmonary pathologies are suspected when the frequency or intensity of respiratory sounds changes or when adventitious sounds, including continuous adventitious sounds (CASs) and discontinuous adventitious sounds (DASs), are identified (Bohadana et al., 2014; Goettel & Herrmann, 2019; Pramono et al., 2017). Patients with coronavirus disease 2019 exhibit adventitious sounds (Wang et al., 2020); hence, auscultation may be a useful approach for disease diagnosis (Raj et al., 2020) and disease progression tracking. However, auscultation performed using a conventional handheld stethoscope involves some limitations (Sovijärvi et al., 1997). First, the interpretation of auscultation results substantially depends on the subjectivity of the practitioners. Even experienced clinicians might not have high consensus rates in their interpretations of auscultatory manifestations (Berry et al., 2016; Grunnreis, 2016). Second, auscultation is a qualitative analysis method. Comparing auscultation results between individuals and quantifying the sound change by reviewing historical records are difficult tasks. Third, prolonged continuous monitoring of respiratory sound is almost impractical.

To overcome the aforementioned limitations, computerized methods for respiratory sound recording and analyses based on traditional signal processing and machine learning have been proposed and reviewed (Gurung et al., 2011; Huq & Moussavi, 2012; Mesaros et al., 2016; Pasterkamp et al., 1997; Pramono et al., 2017). With the advent of the deep learning era, studies have developed novel deep learning–based methods for respiratory sound analysis. However, many of such studies have focused on only distinguishing healthy participants from participants with respiratory disorders (Chambres et al., 2018; Demir et al., 2020; Hosseini et al., 2020; Perna & Tagarelli, 2019; Pham et al., 2020) and distinguishing various types of normal breathing sounds from adventitious sounds (Acharya & Basu, 2020; Aykanat et al., 2017; Bardou et al., 2018; Chen et al., 2019; Grzywalski et al., 2019; Kochetov et al., 2018; Li et al., 2016). Only a few studies (Hsiao et al., 2020; Jácome et al., 2019; Liu et al., 2017; Messner et al., 2018) have explored the use of deep learning for detecting breath phases and adventitious sounds. Moreover, most previous studies on computerized lung sound analysis have been limited by insufficient data. As of writing this paper, the largest reported respiratory sound database is ICBHI 2017 Challenge (Rocha et al., 2017), which comprises 6,898 breath cycles and 10,775 events of wheezes and crackles acquired from 126 individuals.

Data size plays a major role in the creation of a robust and accurate deep learning–based respiratory sound analysis algorithm (Hestness et al., 2017; Sun et al., 2017). Accordingly, the first aim of the present study was to establish a large and open-access respiratory sound database for training such algorithms for the detection of breath phase and adventitious sounds, mainly focusing on lung sounds. The second aim was to conduct a benchmark test on the established lung sound database by using eight recurrent neural network (RNN)-based models. RNNs (Elman, 1990) are effective for time-series analysis; long short-term memory (LSTM; Hochreiter & Schmidhuber, 1997) and gated recurrent unit (GRU; Cho et al., 2014) networks, which are two RNN variants, exhibit superior performance to the original RNN model. However, whether LSTM models are superior to GRU models (and vice versa) in many applications, particularly in respiratory sound analysis, is inconclusive. Bidirectional RNN models (Graves & Schmidhuber, 2005; Schuster & Paliwal, 1997) can transfer not only past information to the future but also future information to the past; these models consistently exhibit superior performance to unidirectional RNN models in many applications (Khandelwal et al., 2016; Linchuan Li et al., 2016; Parascandolo et al., 2016) as well as in breath phase and crackle detection (Messner et al., 2018). However, whether bidirectional RNN models outperform unidirectional RNN models in CAS detection has yet to be determined. Furthermore, the convolutional neural network (CNN)–RNN structure has been proven to be suitable for heart sound analysis (Deng et al., 2020), lung sound analysis (Acharya & Basu, 2020), and other tasks (Linchuan Li et al., 2016; Zhao et al., 2018). Nevertheless, the application of the CNN–RNN structure in respiratory sound detection has yet to be fully investigated. Benchmarking can enable demonstrating the reliability and goodness of a database; it can also be applied to investigate the performance of the RNN variants in respiratory analysis.

In summary, the aims of this study are outlined as follows:

- Establish the largest open-access lung sound database as of writing this paper—HF_Lung_V1 (https://gitlab.com/techsupportHF/HF_Lung_V1).
- Conduct a performance comparison between LSTM and GRU models, between unidirectional and bidirectional models, and between models with and without a CNN in breath phase and adventitious sound detection based on lung sound data.
- Discuss factors influencing model performance.

## 2 Establishment of the lung sound database

### *2.1 Data sources and patients*

The lung sound database was established using two sources. The first source was a database used in a datathon in Taiwan Smart Emergency and Critical Care (TSECC), 2020, under the license of Creative Commons Attribution 4.0 (CC BY 4.0), provided by the Taiwan Society of Emergency and Critical Care Medicine. Lung sound recordings in the TSECC database were acquired from 261 patients.

The second source was sound recordings acquired from 18 residents of a respiratory care ward (RCW) or a respiratory care center (RCC) in Northern Taiwan between August 2018 and October 2019. The recordings were approved by the Research Ethics Review Committee of Far Eastern

Memorial Hospital (case number: 107052-F). This study was conducted in accordance with the 1964 Helsinki Declaration and its later amendments or comparable ethical standards.

All patients were Taiwanese and aged older than 20 years. Descriptive statistics regarding the patients' demographic data, major diagnosis, and comorbidities are presented in Table 1; however, information on the patients in the TSECC database is missing. Moreover, all 18 RCW/RCC residents were under mechanical ventilation.

**Table 1**

Demographic data of patients.

| | Subjects from RCW/RCC (n= 18) | TSECC Database (n = 261) |
|---|---|---|
| Gender (M/F) | 11/7 | NA |
| Age | 67.5 (36.7, 98.3) | NA |
| Height (cm) | 163.6 (147.2, 180.0) | NA |
| Weight (kg) | 62.1 (38.2, 86.1) | NA |
| BMI (kg/m$^2$) | 23.1 (15.6, 30.7) | NA |
| Respiratory Diseases | | |
| ARF | 4 (22.2%) | NA |
| CRF | 8 (44.4%) | NA |
| COPD AE | 1 (5.6%) | NA |
| COPD | 2 (11.1%) | NA |
| Pneumonia | 4 (22.2%) | NA |
| ARDS | 1 (5.6%) | NA |
| Emphysema | 1 (5.6%) | NA |
| Comorbidity | | |
| CKD | 1 (5.6%) | NA |
| AKI | 3 (16.7%) | NA |
| CHF | 2 (11.1%) | NA |
| DM | 7 (38.9%) | NA |
| HTN | 6 (33.3%) | NA |
| Malignancy | 1 (5.6%) | NA |
| Arrythmia | 1 (5.6%) | NA |
| CAD | 1 (5.6%) | NA |

RCW: respiratory care ward, RCC: respiratory care center, ARF: acute respiratory failure, CRF: chronic respiratory failure, COPD AE: chronic obstructive pulmonary disease acute exacerbation, COPD: chronic obstructive pulmonary disease, ARDS: acute respiratory distress syndrome, CKD: chronic kidney disease, AKI: acute kidney injury, CHF: chronic heart failure, DM: diabetes, HTN: hypertension, CAD: cardiovascular disease. The mean values of the age, height, weight, and BMI are presented, with the corresponding 95% CI in parentheses.

## 2.2 *Sound recording*

Breathing lung sounds were recorded using two devices: (1) a commercial electronic stethoscope (Littmann 3200, 3M, Saint Paul, Minnesota, USA) and (2) a customized multichannel acoustic recording device (HF-Type-1) that supports the connection of eight electret microphones. The signals collected by the HF-Type-1 device were transmitted to a tablet (Surface Pro 6, Microsoft, Redmond, Washington, USA; Fig. 1). Breathing lung sounds were collected at the eight locations (denoted by L1–L8) indicated in Fig. 2a. The auscultation locations are described in detail in the caption of Fig. 2. The two devices had a sampling rate of 4,000 Hz and a bit depth of 16 bits. The audio files were recorded in the WAVE (.wav) format.

All lung sounds in the TSECC database were collected using the Littmann 3200 device only, where 15.8-s recordings were obtained sequentially from L1 to L8 (Fig. 2b; Littmann 3200). One round of recording with the Littmann 3200 device entails a recording of lung sounds from L1 to L8. The TSECC database was composed of data obtained from one to three rounds of recording with the Littmann 3200 device for each patient.

We recorded the lung sounds of the 18 RCW/RCC residents by using both the Littmann 3200 device and the HF-Type-1 device. The Littmann 3200 recording protocol was in accordance with that used in the TSECC database, except that data from four to five rounds of lung sound recording were collected instead. The HF-Type-1 device was used to record breath sounds at L1, L2, L4, L5, L6, and L8. One round of recording with the HF-Type-1 device entails a synchronous and continuous recording of breath sounds for 30 min (Fig. 2b; HF-Type-1). However, the recording with the HF-Type-1 device was occasionally interrupted; in this case, the recording duration was <30 min.

Voluntary deep breathing was not mandated during the recording of lung sounds. The statistics of the recordings are listed in Table 2.

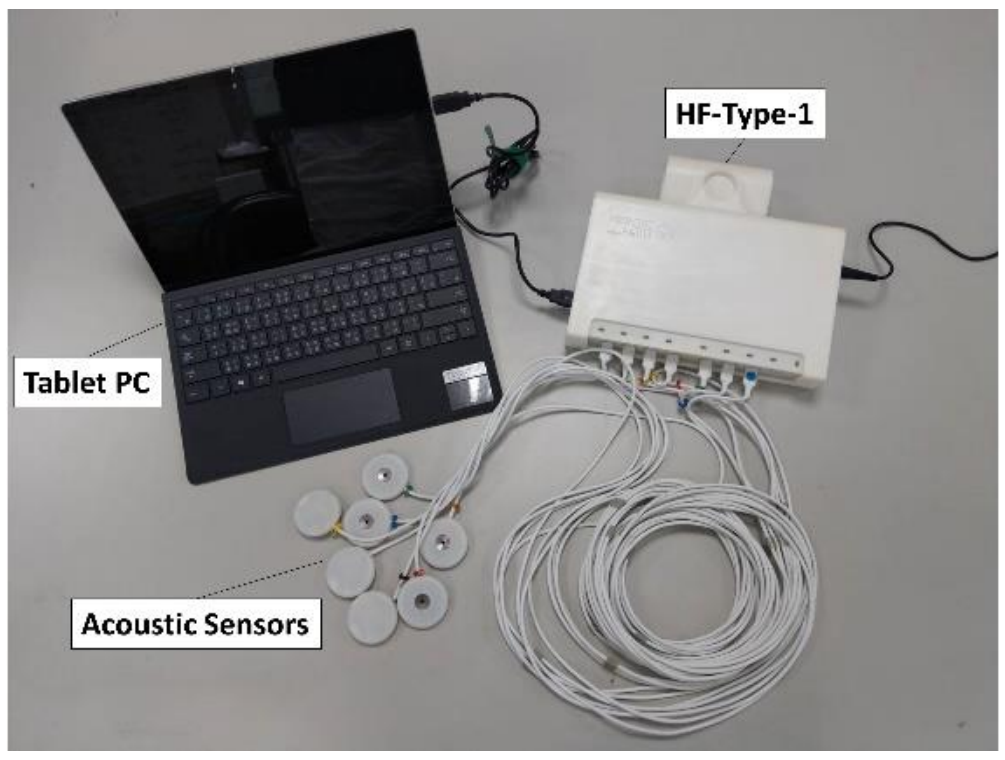

**Fig. 1.** Customized multichannel acoustic recording device (HF-Type-1) connected to a tablet.

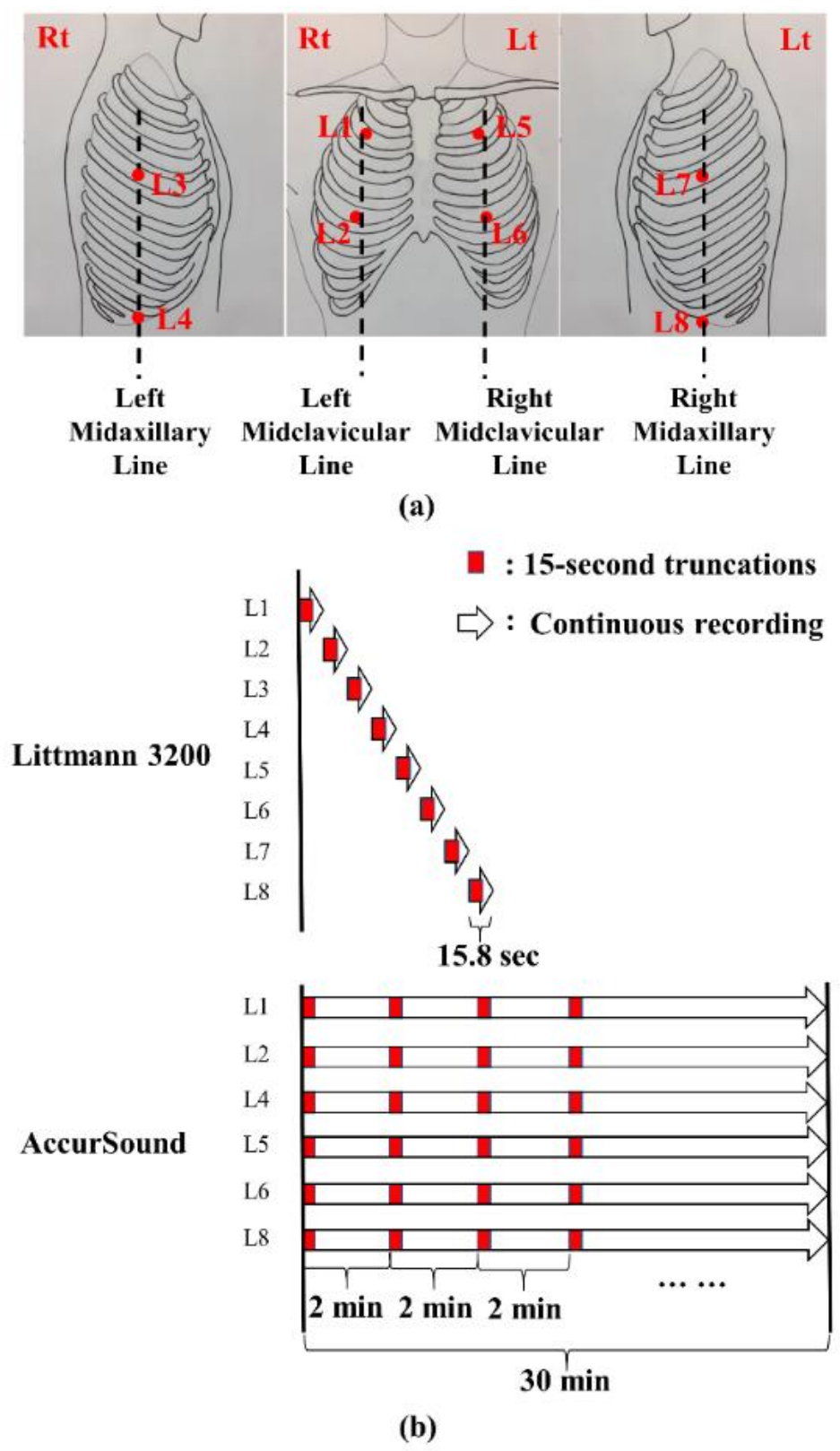

**Fig. 2.** Auscultation locations and lung sound recording protocol. (a) Auscultation locations (L1–L8): L1: second intercostal space (ICS) on the right midclavicular line (MCL); L2: fifth ICS on the right MCL; L3: fourth ICS on the right midaxillary line (MAL); L4: tenth ICS on the right MAL; L5: second ICS on the left MCL; L6: fifth ICS on the left MCL; L7: fourth ICS on the left MAL; and L8: tenth ICS on the left MAL. (b) A standard round of breathing lung sound recording with Littmann 3200 and HF-Type-1 devices. The white arrows represent a continuous recording, and the small red blocks represent 15-s recordings. When the Littmann 3200 device was used, 15.8-s signals were recorded sequentially from L1 to L8. Subsequently, all recordings were truncated to 15 s. When the HF-Type-1 device was used, sounds at L1, L2, L4, L5, L6, and L8 were recorded simultaneously. Subsequently, each 2-min signal was truncated to generate new 15-s audio files.

**Table 2**

Statistics of recordings and labels of HF_Lung_V1 database.

| | Littmann 3200 | HF-Type-1 | Total |
|---|---|---|---|
| Subjects | | | |
| n | 261 | 18 | 261 |
| Recordings | | | |
| Filename prefix | steth_ | trunc_ | NA |
| Rounds of recording | 748 | 70 | NA |
| No of 15-sec recordings | 4504 | 5261 | 9765 |
| Total duration (min) | 1126 | 1315.25 | 2441.25 |
| Labels | | | |
| No of I | 16535 | 17560 | 34095 |
| Total duration of I (min) | 257.17 | 271.02 | 528.19 |
| Mean duration of I (sec) | 0.93 | 0.93 | 0.93 |
| No of E | 9107 | 9242 | 18349 |
| Total duration of E (min) | 160.25 | 132.60 | 292.85 |
| Mean duration of E (sec) | 1.06 | 0.86 | 0.96 |
| No of C/W/S/R | 6984/3974/152/2858 | 6899/4483/534/1882 | 13883/8457/686/4740 |
| Total duration of C/W/S/R (min) | 105.90/63.92/1.94/40.04 | 85.26/55.80/7.52/21.94 | 191.16/119.73/9.46/61.98 |
| Mean duration of C/W/S/R (sec) | 0.91/0.97/0.76/0.84 | 0.74/0.75/0.85/0.70 | 0.83/0.85/0.83/0.78 |
| No of D | 7266 | 8340 | 15606 |
| Total duration of D (min) | 111.75 | 55.80 | 230.87 |
| Mean duration of D (sec) | 0.92 | 0.87 | 0.89 |

I: inhalation, E: exhalation, W: wheeze, S: stridor, R: rhonchus, C: continuous adventitious sound, D: discontinuous adventitious sound. W, S, and R were combined to form C.

### *2.3 Audio file truncation*

In this study, the standard duration of an audio signal used for inhalation, exhalation, and adventitious sound detection was 15 s. This duration was selected because a 15-s signal contains at least three complete breath cycles, which are adequate for a clinician to reach a clinical conclusion.

Furthermore, a 15-s breath sound was be used previously for verification and validation (Pasterkamp et al., 2016) .

Because each audio file generated by the Littmann 3200 device had a length of 15.8 s, we cropped out the final 0.8-s signal from the files (Fig. 2b; Littmann 3200). Moreover, we used only the first 15 s of each 2-min signal of the audio files (Fig. 2b; HF-Type-1) generated by the HF-Type-1 device. Table 2 presents the number of truncated 15-s recordings and the total duration.

### *2.4 Data labeling*

Because the data in the TSECC database contains only classification labels indicating whether a CAS or DAS exists in a recording, we attempted to label the event level of all sound recordings. Two board-certified respiratory therapists (NJL and YLW) and one board-certified nurse (WLT), with 8, 3, and 13 years of clinical experience, respectively, were recruited to label the start and end points of inhalation (I), exhalation (E), wheeze (W), stridor (S), rhonchus (R), and DAS (D) events in the recordings. They labeled the sound events by listening to the recorded breath sounds while simultaneously observing the corresponding patterns on a spectrogram by using customized labeling software (Hsu et al., 2021). The labelers were asked not to label sound events if they could not clearly identify the corresponding sound or if an incomplete event at the beginning or end of an audio file caused difficulty in identification. BFH held regular meetings to ensure that the labelers had good agreement on labeling criteria based on a few samples by judging the mean pseudo-κ value (Jácome et al., 2019). When developing artificial intelligence (AI) detection models, we combined the W, S, and R labels to form CAS labels (C). Moreover, the D labels comprised only crackles, which were not differentiated into coarse or fine crackles. The labelers were asked to label the period containing crackles but not a single explosive sound (generally less than 25 ms) of a crackle. Each recording was annotated by only one labeler; thus, the labels did not represent perfect ground truth. However, we used the labels as ground-truth labels for model training, validation, and testing. The statistics of the labels are listed in Table 2.

## 3. Inhalation, exhalation, CAS, and DAS detection

### *3.1 Framework*

The inhalation, exhalation, CAS, and DAS detection framework developed in this study is displayed in Fig. 3. The prominent advantage of the research framework is its modular design. Specifically, each unit of the framework can be tested separately, and the algorithms in different parts of the framework can be modified to achieve optimal overall performance. Moreover, the output of some blocks can be used for multiple purposes. For instance, the spectrogram generated by the preprocessing block can be used as the input of a model or for visualization in the user interface for real-time monitoring.

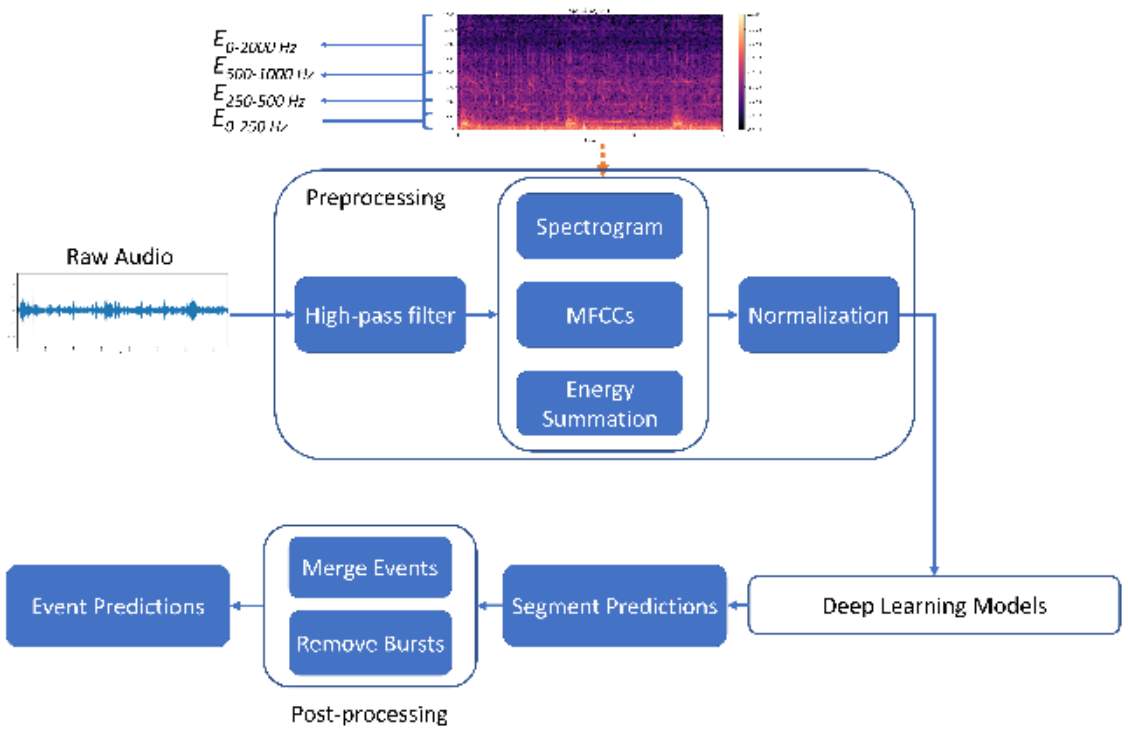

**Fig. 3.** Pipeline of detection framework.

The framework comprises three parts: preprocessing, deep learning–based modeling, and postprocessing. The preprocessing part involves signal processing and feature engineering techniques. The deep learning–based modeling part entails the use of a well-designed neural network for obtaining a sequence of classification predictions rather than a single prediction. The postprocessing part involves merging the segment prediction results and eliminating the burst event.

### *3.2 Preprocessing*

We processed the lung sound recordings at a sampling frequency of 4 kHz. First, to eliminate the 60-Hz electrical interference and a part of the heart sound noise, we applied a high-pass filter to the recordings by setting a filter order of 10 and cut-off frequency of 80 Hz. The filtered signals were then processed using the short-time Fourier transform (STFT). In the STFT, we set a Hanning window size of 256 and hop length of 64; no additional zero-padding was applied. Thus, a 15-s sound signal could be transformed into a corresponding spectrogram with a size of 938 × 129. To obtain the spectral information regarding the lung sounds, we extracted the following features (Chamberlain et al., 2016; Messner et al., 2018):

- Spectrogram: We extracted 129-bin log-magnitude spectrograms.
- Mel frequency cepstral coefficients (MFCCs): We extracted 20 static coefficients, 20 delta coefficients (Δ), and 20 acceleration coefficients ($\Delta^2$). We used 40 mel bands within a frequency range of 0–4,000 Hz. The frame width used to calculate the delta and acceleration coefficients was set to 9, which resulted in a 60-bin vector per frame.
- Energy summation: We computed the energy summation of four frequency bands, namely 0–250, 250–500, 500–1,000, and 0–2,000 Hz, and obtained four values per time frame.

After extracting the aforementioned features, we concatenated them to form a 938 × 193 feature matrix. Subsequently, we conducted min–max normalization on each feature. The values of the normalized features ranged between 0 and 1.

### *3.3 Deep learning models*

We investigated the performance of eight RNN models, namely LSTM, GRU, bidirectional LSTM (BiLSTM), bidirectional GRU (BiGRU), CNN-LSTM, CNN-GRU, CNN-BiLSTM, and CNN-BiGRU, in terms of inhalation, exhalation, and adventitious sound detection. Fig. 4 illustrates the detailed model structures. The outputs of the LSTM, GRU, BiLSTM, and BiGRU models were 938 × 1 vectors, and those of the CNN-LSTM, CNN-GRU, CNN-BiLSTM, and CNN-BiGRU models were 469 × 1 vectors. An element in these vectors was set to 1 if an inhalation, exhalation, CAS, or DAS occurred within a time segment in which the output value passed the thresholding criterion; otherwise, the element was set to 0.

For a fairer comparison of the performance of the unidirectional and bidirectional models, we trained additional simplified (SIMP) BiLSTM, SIMP BiGRU, SIMP CNN-BiLSTM, and SIMP CNN-BiGRU models by adjusting the number of trainable parameters. Parameter adjustment was conducted by halving the number of cells of the LSTM and GRU layers.

We used Adam as the optimizer in the benchmark model, and we set the initial learning rate to 0.0001 with a step decay (0.2×) when the validation loss did not decrease for 10 epochs. The learning process stopped when no improvement occurred over 50 consecutive epochs.

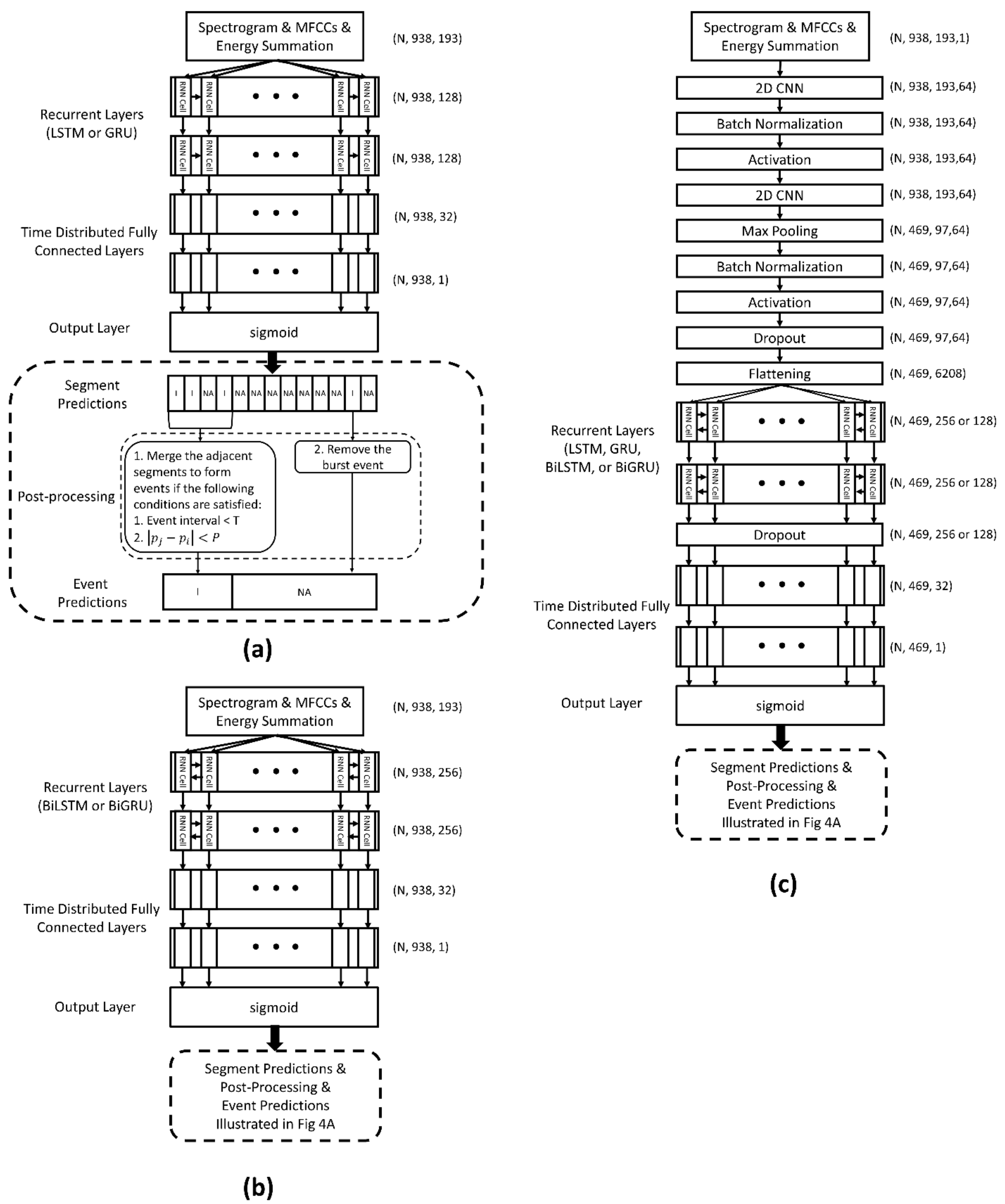

**Fig. 4.** Model architectures and postprocessing for inhalation, exhalation, CAS, and DAS segment and event detection. (a) LSTM and GRU models; (b) BiLSTM and BiGRU models; and (c) CNN-LSTM, CNN-GRU, CNN-BiLSTM, and CNN-BiGRU models.

### 3.4 Postprocessing

The prediction vectors obtained using the adopted models can be further processed for different purposes. For example, we can transform the prediction result from frames to time for real-time monitoring. The breathing duration of most humans lies within a certain range; we considered this fact in our study. Accordingly, when the prediction results obtained using the models indicated that two consecutive inhalation events occurred within a very small interval, we checked the continuity of these two events and decided whether to merge them, as illustrated in the bottom panel of Fig. 4a. For example, when the interval between the $j$th and $i$th events was smaller than $T$ s, we computed the difference in frequency between their energy peaks ($|\boldsymbol{p_j} - \boldsymbol{p_i}|$). Subsequently, if the difference was below a given threshold $P$, the two events were merged into a single event. In the experiment, $T$ was set to 0.5 s, and $P$ was set to 25 Hz. After the merging process, we further assessed whether a burst event existed. If the duration of an event was shorter than 0.05 s, the event was deleted.

### 3.5 Dataset arrangement and cross-validation

We adopted fivefold cross-validation in the training dataset to train and validate the models. Moreover, we used an independent testing dataset to test the performance of the trained models. According to our preliminary experience, the acoustic patterns of the breath sounds collected from one patient at different auscultation locations or between short intervals had many similarities. To avoid potential data leakage caused by our methods of collecting and truncating the breath sound signals, we assigned all truncated recordings collected on the same day to only one of the training, validation, or testing datasets; this is because these recordings might have been collected from the same patient within a short period. The statistics of the datasets are listed in Table 3. We used only audio files containing CASs and DASs to train and test their corresponding detection models.

**Table 3**

Statistics of the datasets and labels of the HF_Lung_V1 database.

| | Training Dataset | Testing Dataset | Total |
|---|---|---|---|
| Recordings | | | |
| No of 15-sec recordings | 7809 | 1956 | 9765 |
| Total duration (min) | 1952.25 | 489 | 2441.25 |
| Labels | | | |
| No of I | 27223 | 6872 | 34095 |
| Total duration of I (min) | 422.17 | 105.97 | 528.14 |
| Mean duration of I (sec) | 0.93 | 0.93 | 0.93 |
| No of E | 15601 | 2748 | 18349 |
| Total duration of E (min) | 248.05 | 44.81 | 292.85 |
| Mean duration of E (sec) | 0.95 | 0.98 | 0.96 |
| No of C/W/S/R | 11464/7027/657/3780 | 2419/1430/29/960 | 13883/8457/686/4740 |
| Total duration of C/W/S/R (min) | 160.16/100.71/9.10/50.35 | 31.01/19.02/0.36/11.63 | 191.16/119.73/9.46/61.98 |
| Mean duration of C/W/S/R (sec) | 0.84/0.86/0.83/0.80 | 0.77/0.80/0.74/0.73 | 0.83/0.85/0.83/0.78 |
| No of D | 13794 | 1812 | 15606 |
| Total duration of D (min) | 203.59 | 27.29 | 230.87 |
| Mean duration of D (sec) | 0.89 | 0.90 | 0.89 |

I: inhalation, E: exhalation, W: wheeze, S: stridor, R: rhonchus, C: continuous adventitious sound, D: discontinuous adventitious sound. W, S, and R were combined to form C.

### *3.6 Task definition and evaluation metrics*

Pramono et al. (2017) clearly defined classification and detection at the segment, event, and recording levels. In this study, we performed two tasks. The first task involved performing detection at the segment level. The acoustic signal of each lung sound recording was transformed into a spectrogram. The temporal resolution of the spectrogram depended on the window size and overlap ratio of the STFT. The aforementioned parameters were fixed such that each spectrogram was a matrix of size 938 × 129. Thus, each recording contained 938 time segments (time frames), and each time segment was automatically labeled (Fig. 5b) according to the ground-truth event labels (Fig. 5a) assigned by the labelers. The output of the prediction process was a sequential prediction matrix (Fig. 5c) of size 938 × 1 in the LSTM, GRU, BiLSTM, and BiGRU models and size 469 × 1 in the CNN-LSTM, CNN-GRU, CNN-BiLSTM, and CNN-BiGRU models. By comparing the sequential prediction with the ground-truth time segments, we could define true positive (TP; orange vertical bars in Fig. 5d), true negative (TN; green vertical bars in Fig. 5d), false positive (FP; black vertical bars in Fig. 5d), and false negative (FN; yellow vertical bars in Fig. 5d) time segments. Subsequently, the models' sensitivity and specificity in classifying the segments in each recording were computed.

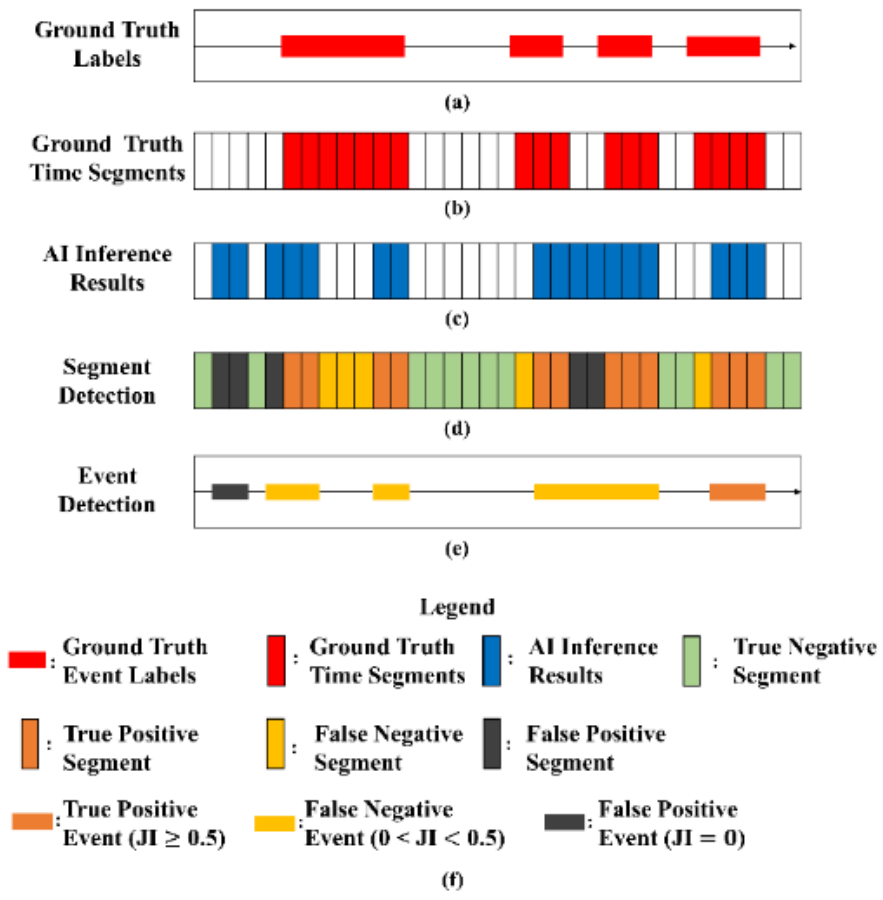

**Fig. 5.** Task definition and evaluation metrics. (a) Ground-truth event labels, (b) ground-truth time segments, (c) AI inference results, (d) segment classification, (e) event detection, and (f) legend. JI: Jaccard index.

The second task entailed event detection at the recording level. After completing the sequential prediction (Fig. 5c), we assembled the time segments associated with the same label into a corresponding event (Fig. 5e). We also derived the start and end times of each assembled event. The Jaccard index (JI; Jácome et al., 2019) was used to determine whether an AI inference result correctly matched the ground-truth event. For an assembled event to be designated as a TP event (orange horizontal bars in Fig. 5e), the corresponding JI value must be greater than 0.5. If the JI was between 0 and 0.5, the assembled event was designated as an FN event (yellow horizontal bars in Fig. 5e), and if it was 0, the assembled event was designated as an FP event (black horizontal bars in Fig. 5e). A TN event cannot be defined in the task of event detection.

The performance of the models was evaluated using the *F1* score, and that of segment detection was evaluated using the receiver operating characteristic (ROC) curve and area under the ROC curve (AUC). In addition, the mean absolute percentage error (MAPE) of event detection was derived. The accuracy, positive predictive value (PPV), sensitivity, specificity, and *F1* score of the models are presented in Appendix A.

### *3.7 Hardware and software*

We trained the baseline models on an Ubuntu 18.04 server that was provided by the National Center for High-Performance Computing in Taiwan [Taiwan Computing Cloud (TWCC)] and was equipped with an Intel(R) Xeon(R) Gold 6154 @3.00 GHz CPU with 90 GB RAM. To manage the intensive computation involved in RNN training, we implemented the training module by using the TensorFlow 2.10, CUDA 10, and CuDNN 7 programs to run the NVIDIA Titan V100 card on the TWCC server for GPU acceleration.

## 4 Results

### *4.1 LSTM versus GRU models*

Table 4 presents the *F1* scores used to compare the eight LSTM- and GRU-based models. When a CNN was not added, the GRU models outperformed the LSTM models by 0.7%–9.5% in terms of the *F1* scores. However, the CNN-GRU and CNN-BiGRU models did not outperform the CNN-LSTM and CNN-BiLSTM models in terms of the *F1* scores (and vice versa).

According to the ROC curves presented in Fig. 6a–d, the GRU-based models outperformed the LSTM-based models in all compared pairs, except for one pair, in terms of DAS segment detection (AUC of 0.891 for CNN-BiLSTM vs 0.889 for CNN-BiGRU).

**Table 4**
Comparison of *F1* scores between LSTM-based models and GRU-based models.

| | | Inhalation | | Exhalation | | CASs | | DASs | |
|---|---|---|---|---|---|---|---|---|---|
| | n of | *F1* score | | *F1* score | | *F1* score | | *F1* score | |
| Models | trainable parameters | Segment Detection | Event Detection | Segment Detection | Event Detection | Segment Detection | Event Detection | Segment Detection | Event Detection |
| LSTM | 300,609 | 73.9% | 76.1% | 51.8% | 57.0% | 15.1% | 12.2% | 62.6% | 59.1% |
| GRU | 227,265 | **76.2%** | **78.9%** | **59.8%** | **65.6%** | **24.6%** | **20.1%** | **65.9%** | **62.5%** |
| BiLSTM | 732,225 | 78.1% | 84.0% | 57.3% | 63.9% | 19.8% | 19.1% | 69.6% | 70.0% |
| BiGRU | 552,769 | **80.3%** | **86.2%** | **64.1%** | **70.9%** | **26.9%** | **25.6%** | **70.3%** | **71.4%** |
| CNN-LSTM | 3,448,513 | 77.6% | 81.1% | **57.7%** | **62.1%** | 45.3% | 42.5% | **68.8%** | 64.4% |
| CNN-GRU | 2,605,249 | **78.4%** | **82.0%** | 57.2% | 62.0% | **51.5%** | **49.8%** | 68.0% | **64.6%** |
| CNN-BiLSTM | 6,959,809 | **80.6%** | **86.3%** | 60.4% | 65.6% | 47.9% | 46.4% | **71.2%** | **70.8%** |
| CNN-BiGRU | 5,240,513 | **80.6%** | 86.2% | **62.2%** | **68.5%** | **53.3%** | **51.6%** | 70.6% | 70.0% |

The bold values indicate the higher *F1* score between the compared pairs of models.

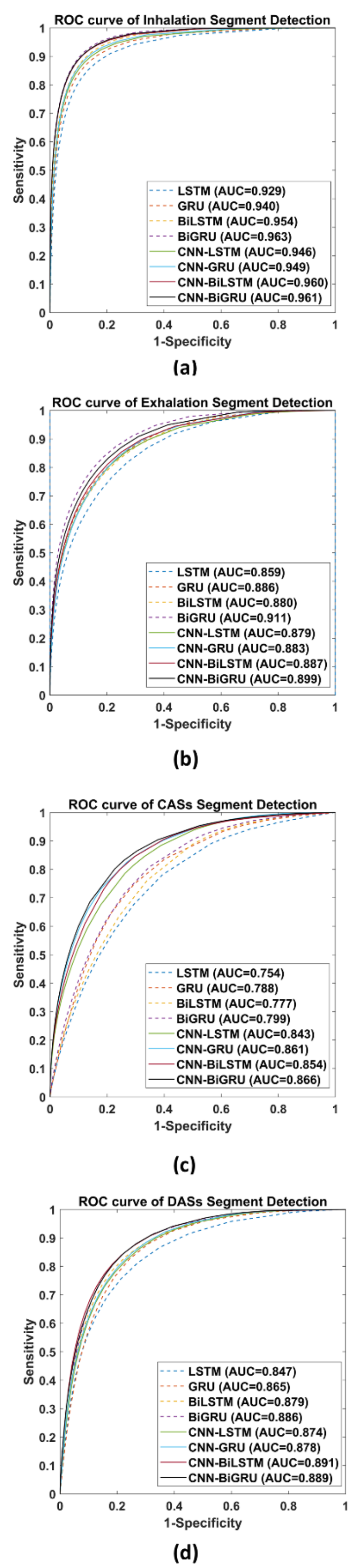

**Fig. 6.** ROC curves for (a) inhalation, (b) exhalation, (c) CAS, and (d) DAS segment detection. The corresponding AUC values are presented.

*4.2 Unidirectional versus bidirectional models*

**Table 5**

Comparison of *F1* scores between the unidirectional and bidirectional models.

| Models | n of trainable parameters | Inhalation | | Exhalation | | CASs | | DASs | |
|---|---|---|---|---|---|---|---|---|---|
| | | *F1* score | | *F1* score | | *F1* score | | *F1* score | |
| | | Segment Detection | Event Detection | Segment Detection | Event Detection | Segment Detection | Event Detection | Segment Detection | Event Detection |
| LSTM | 300,609 | 73.9% | 76.1% | 51.8% | 57.0% | 15.1% | 12.2% | 62.6% | 59.1% |
| SIMP BiLSTM | 235,073 | **77.8%** | **84.1%** | **55.8%** | **62.4%** | **19.8%** | **17.9%** | **68.8%** | **68.9%** |
| GRU | 227,265 | 76.2% | 78.9% | 59.8% | 65.6% | 24.6% | 20.1% | 65.9% | 62.5% |
| SIMP BiGRU | 178,113 | **80.1%** | **86.1%** | **63.7%** | **70.0%** | **25.0%** | **22.2%** | **70.3%** | **71.3%** |
| CNN-LSTM | 3,448,513 | 77.6% | 81.1% | 57.7% | 62.1% | 45.3% | 42.5% | 68.8% | 64.4% |
| SIMP CNN-BiLSTM | 3,382,977 | **80.0%** | **85.8%** | **60.4%** | **66.2%** | **50.8%** | **50.2%** | **70.2%** | **70.2%** |
| CNN-GRU | 2,605,249 | 78.4% | 82.0% | 57.2% | 62.0% | 51.5% | 49.8% | 68.0% | 64.6% |
| SIMP CNN-BiGRU | 2,556,097 | **80.1%** | **85.9%** | **62.4%** | **68.4%** | **52.6%** | **51.5%** | **69.9%** | **69.5%** |

The bold values indicate the higher *F1* score between the compared pairs of models. SIMP means the number of trainable parameters is adjusted.

As presented in Table 5, the bidirectional models outperformed their unidirectional counterparts in all the defined tasks by 0.4%–9.8% in terms of the *F1* scores, even when the bidirectional models had fewer trainable parameters after model adjustment.

*4.3 Models with CNN versus those without CNN*

According to Table 6, the models with a CNN outperformed those without a CNN in 26 of the 32 compared pairs.

The models with a CNN exhibited higher AUC values than did those without a CNN (Fig. 6a–d),

except that BiGRU had a higher AUC value than did CNN-BiGRU in terms of inhalation detection (0.963 vs 0.961), GRU had a higher AUC value than did CNN-GRU in terms of exhalation detection (0.886 vs 0.883), and BiGRU had a higher AUC value than did CNN-BiGRU in terms of exhalation detection (0.911 vs 0.899).

Moreover, compared with the LSTM, GRU, BiLSTM, and BiGRU models, the CNN-LSTM, CNN-GRU, CNN-BiLSTM, and CNN-BiGRU models exhibited flatter and lower MAPE curves over a wide range of threshold values in all event detection tasks (Fig. 7a–d).

**Table 6**

Comparison of *F1* scores between models without and with a CNN.

| | | Inhalation | | Exhalation | | CASs | | DASs | |
|---|---|---|---|---|---|---|---|---|---|
| | n of trainable parameters | *F1* score | | *F1* score | | *F1* score | | *F1* score | |
| Models | | Segment Detection | Event Detection | Segment Detection | Event Detection | Segment Detection | Event Detection | Segment Detection | Event Detection |
| LSTM | 300,609 | 73.9% | 76.1% | 51.8% | 57.0% | 15.10% | 12.20% | 62.60% | 59.10% |
| CNN-LSTM | 3,448,513 | **77.6%** | **81.1%** | **57.7%** | **62.1%** | **45.30%** | **42.50%** | **68.80%** | **64.40%** |
| BiLSTM | 732,225 | 76.2% | 78.9% | **59.8%** | **65.6%** | 19.80% | 17.90% | 68.80% | 68.90% |
| CNN-BiLSTM | 6,959,809 | **78.4%** | **82.0%** | 57.2% | 62.0% | **50.80%** | **50.20%** | **70.20%** | **70.20%** |
| GRU | 227,265 | 78.1% | 84.0% | 57.3% | 63.9% | 24.60% | 20.10% | 65.90% | 62.50% |
| CNN-GRU | 2,605,249 | **80.6%** | **86.3%** | **60.4%** | **65.6%** | **51.50%** | **49.80%** | **68.00%** | **64.60%** |
| BiGRU | 178,113 | 80.3% | **86.2%** | **64.1%** | **70.9%** | 25.00% | 22.20% | **70.30%** | **71.30%** |
| CNN-BiGRU | 2,556,097 | **80.6%** | **86.2%** | 62.2% | 68.5% | **52.60%** | **51.50%** | 69.90% | 69.50% |

The bold values indicate the higher *F1* score between the compared pairs of models.

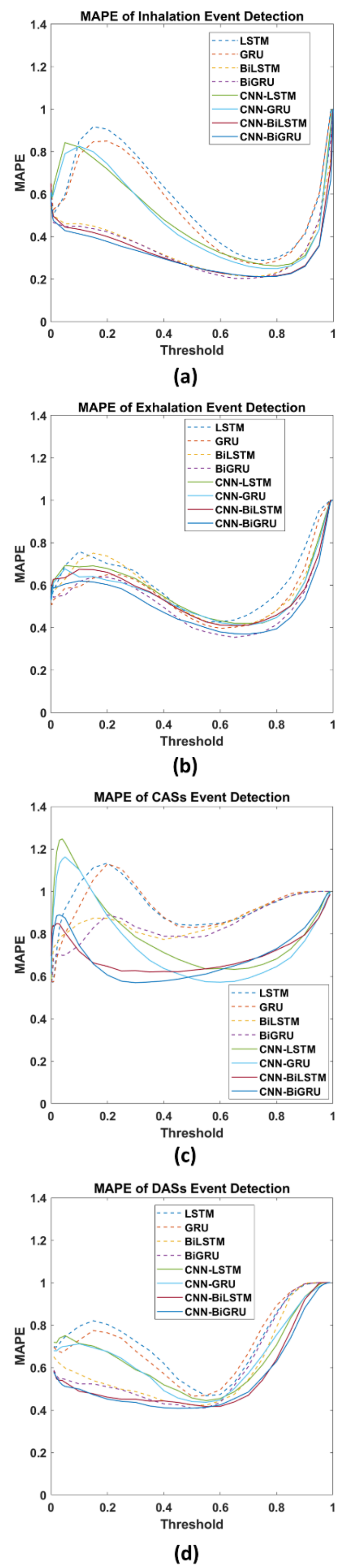

**Fig. 7.** MAPE curves for (a) inhalation, (b) exhalation, (c) CAS, and (d) DAS event detection.

# 5 Discussion

## *5.1 Benchmark results*

According to the *F1* scores presented in Table 4, among models without a CNN, the GRU and BiGRU models consistently outperformed the LSTM and BiLSTM models in all defined tasks. However, the GRU-based models did not have superior *F1* scores among models with a CNN. Regarding the ROC curves and AUC values (Fig. 6a–d), the GRU-based models consistently outperformed the other models in all but one task. Accordingly, we can conclude that GRU-based models perform slightly better than LSTM-based models in lung sound analysis. Previous studies have also compared LSTM- and GRU-based models (Chung et al., 2014; Khandelwal et al., 2016; Shewalkar, 2018). Although a concrete conclusion cannot be drawn regarding whether LSTM-based models are superior to the GRU-based models (and vice versa), GRU-based models have been reported to outperform LSTM-based models in terms of computation time (Khandelwal et al., 2016; Shewalkar, 2018).

As presented in Table 5, the bidirectional models outperformed their unidirectional counterparts in all defined tasks, a finding that is consistent with several previously obtained results (Graves & Schmidhuber, 2005; Khandelwal et al., 2016; Messner et al., 2018; Parascandolo et al., 2016).

A CNN can facilitate the extraction of useful features and enhance the prediction accuracy of RNN-based models. The benefits engendered by a CNN are particularly vital in CAS detection. For the models with a CNN, the *F1* score improvement ranged from 26.0% to 30.3% and the AUC improvement ranged from 0.067 to 0.089 in the CAS detection tasks. Accordingly, we can infer that considerable information used in CAS detection resides in the local positional arrangement of the features. Thus, a two-dimensional CNN facilitates the extraction of the associated information. Notably, CNN-induced improvements in model performance in the inhalation, exhalation, and DAS detection tasks were not as high as those observed in the CAS detection tasks. The MAPE curves (Fig. 7a–d) reveal that a model with a CNN has more consistent predictions over various threshold values.

In our previous study (Hsiao et al., 2020), an attention-based encoder–decoder architecture based on ResNet and LSTM exhibited favorable performance in inhalation (*F1* score of 90.4%) and exhalation (*F1* score of 93.2%) segment detection tasks. However, the model was established on the basis of a very small dataset (489 recordings of 15-s-long lung sounds). Moreover, the model involves a complicated architecture; hence, it is impossible to implement real-time respiratory monitoring in devices with limited computing power, such as smartphones or medical-grade tablets.

Few studies have performed event detection at the recording level by using a comparatively simple deep learning model. Messner et al. (2018) used the BiGRU model and one-dimensional labels (similar to those used in the present study) for breath phase and crackle detection. Their BiGRU model exhibited comparable performance to our models in terms of inhalation event detection (*F1* scores, 87.0% vs 86.2%) and in terms of DAS event detection (*F1* scores, 72.1% vs 71.4%). However, the performance of the BiGRU model differed considerably from that of our models in terms of exhalation detection (*F1* scores: 84.6% vs 70.9%). One of the reasons for this discrepancy is that Messner et al. (2018) established their ground-truth labels on the basis of the gold-standard signals of a pneumotachograph. Another reason is that an exhalation label is not always available following an inhalation label in our data. Finally, we did not specifically control the sounds we recorded; for example, we did not ask patients to perform voluntary deep breathing or keep ambient noise down. The factors influencing the model performance are further discussed in the next section.

### *5.2 Factors influencing model performance*

The benchmark performance of the proposed models may have been influenced by the following factors: (1) unusual breathing patterns; (2) imbalanced data; (3) low signal-to-noise ratio (SNR); (4) noisy labels, including class and attribute noise, in the database; and (5) sound overlapping.

Fig. 8 displays most of the breath patterns present in the HF_Lung_V1 database. Fig. 8a illustrates the general pattern of a breath cycle in the lung sounds when the ratio of inhalation to exhalation durations is approximately 2:1 and an expiratory pause is noted (Pramono et al., 2017; Sarkar et al., 2015). Fig. 8b presents a frequent condition under which an exhalation is not completely heard by the labelers. However, because we did not ask the subjects to breath voluntarily when recording the sound, many unusual breath patterns might have been recorded, such as patterns caused by shallow breathing, fast breathing, and apnea as well as those caused by double triggering of the ventilator (Thille et al., 2006) and air trapping (Blanch et al., 2005; Miller et al., 2014). These unusual breathing patterns might confuse the labeling and learning processes and result in poor testing results.

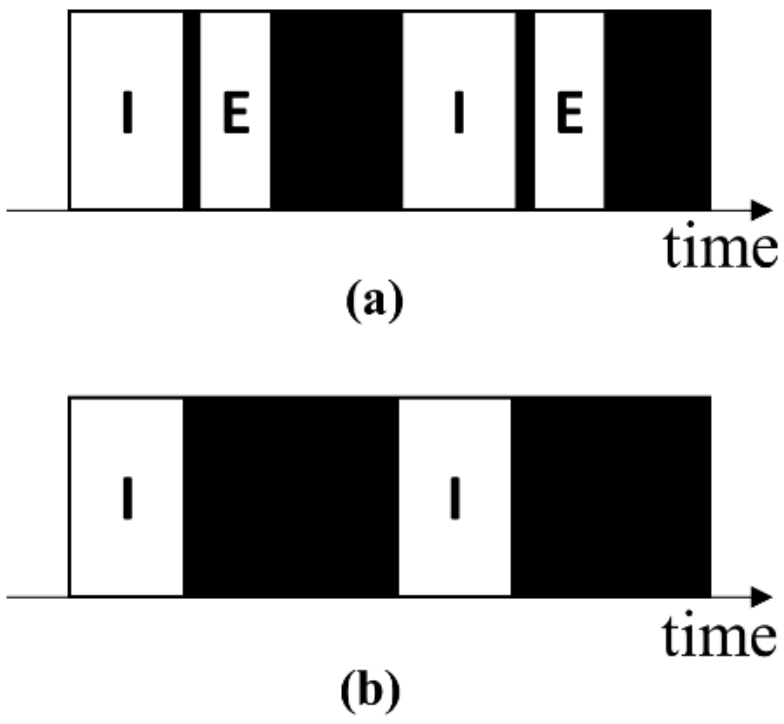

**Fig. 8.** Patterns of normal breathing lung sounds. (a) General lung sound patterns and (b) general lung sound patterns with unidentifiable exhalations. “I” represents an identifiable inhalation event, “E” represents an identifiable exhalation event, and the black areas represent pause phases.

The developed database contains imbalanced numbers of inhalation and exhalation labels (34,095 and 18,349, respectively) because not every exhalation was heard and labeled. In addition, the proposed models may possess the capability of learning the rhythmic rise and fall of breathing signals but not the capability of learning acoustic or texture features that can distinguish an inhalation from an exhalation. This may thus explain the models’ poor performance in exhalation detection. However, these models are suitable for respiratory rate estimation and apnea detection as long as appropriate inhalation detection is achieved. Furthermore, for all labels, the summation of the event duration was smaller than that of the background signal duration (these factors had a ratio of approximately 1:2.5 to 1:7). The aforementioned phenomenon can be regarded as foreground–background class imbalance (Oksuz et al., 2020) and will be addressed in future studies.

Most of the sounds in the established database were not recorded during the patients performed deep breathing; thus, the signal quality was not maximized. However, training models with such nonoptimal data increase their adaptability to real-world scenarios. Moreover, the SNR may be reduced by noise, such as human voices; music; sounds from bedside monitors, televisions, air conditioners, fans, and radios; sounds generated by mechanical ventilators; electrical noise generated by touching or moving the parts of acoustic sensors; and friction sounds generated by the rubbing of two surfaces together (e.g., rubbing clothes with the skin). A poor SNR of audio signals can lead to difficulties in labeling and prediction tasks. The features of some noise types are considerably similar to those of adventitious sounds. The poor performance of the proposed models in CAS detection can be partly attributed to the noisy environment in which the lung sounds were recorded. In particular, the sounds generated by ventilators caused numerous FP events in the CAS detection tasks. Thus, additional effort is required to develop a superior preprocessing algorithm that can filter out influential noise or to identify a strategy to ensure that models focus on learning the correct CAS features. Furthermore, the integration of active noise-canceling technology (Wu et al., 2020) or noise suppression technology (Emmanouilidou et al., 2017) into respiratory sound monitors can help reduce the noise from auscultatory signals.

The sound recordings in the HF_Lung_V1 database were labeled by only one labeler; thus,

some noisy labels, including class and attribute noise, may exist in the database (Zhu & Wu, 2004). These noisy labels are attributable to (1) the different hearing abilities of the labeler, which can cause differences in the labeled duration; (2) the absence of clear criteria for differentiating between target and confusing events; (3) individual human errors; (4) tendency to not label events located close to the beginning and end of a recording; and (5) confusion caused by unusual breath patterns and poor SNRs. However, deep learning models exhibit high robustness to noisy labels (Rolnick et al., 2017). Accordingly, we are currently working toward establishing better ground-truth labels.

Breathing generates CASs and DASs under abnormal respiratory conditions. This means that the breathing sound, CAS, and DAS might overlap with one another during the same period. This sound overlapping, along with the data imbalance, makes the CAS and DAS detection models learn to read the rise and fall of the breathing energy and falsely identify an inhalation or exhalation as CAS or DAS, respectively. This FP detection was observed in our benchmark results. In the future, strategies must be adopted to address the problem of sound overlap.

## 6 Conclusion

We established a large open-access lung sound database, namely HF_Lung_V1 (https://gitlab.com/techsupportHF/HF_Lung_V1), that contains 9,765 audio files of lung sounds (each with a duration of 15 s), 34,095 inhalation labels, 18,349 exhalation labels, 13,883 CAS labels (comprising 8,457 wheeze labels, 686 stridor labels, and 4,740 rhonchus labels), and 15,606 DAS labels (all of which are crackles).

We also investigated the performance of eight RNN-based models in terms of inhalation, exhalation, CAS detection, and DAS detection in the HF_Lung_V1 database. We determined that the bidirectional models outperformed the unidirectional models in lung sound analysis. Furthermore, the addition of a CNN to these models further improved their performance.

Future studies can develop more accurate respiratory sound analysis models. First, highly accurate ground-truth labels should be established. Second, researchers should investigate the performance of RNN-based models containing state-of-the-art convolutional layers. Third, regional CNN variants can be adopted in lung sound analysis if the labels are expanded to two-dimensional bounding boxes (Jácome et al., 2019). Fourth, wavelet-based approaches, empirical mode decomposition, and other methods that can extract different features should be investigated (Pramono et al., 2017; Pramono et al., 2019). Finally, respiratory sound monitors should be equipped with the capability of tracheal breath sound analysis (Wu et al., 2020).

## Author Contributions

YCL, NJL, YLW, and BFH collected the breath sounds. NJL, YLW, and WLT established the labels. BFH and ZLT organized the labels. JH and CWC helped recruit the study participants. FSH, SRH, YRC, and FPL conceptualized the research. CWH and CCC trained the deep learning models. FSH, SRH, LCC, YTC, and CTT contributed to the final manuscript. FPL supervised the research.

## Acknowledgments

This study was partially funded by the Raising Children Medical Foundation, Taiwan. The authors thank the employees of Heroic Faith Medical Science Co. Ltd. who have ever partially contributed to developing the HF-Type-1 and establishing the HF_Lung_V1 database. This manuscript was edited by Wallace Academic Editing. We also thank the All Vista Healthcare Center, Ministry of Science and Technology, Taiwan for the support.

## Conflicts of Interest

The authors declare that they have no conflicts of interest relevant to this research.

## Appendix A

Appendix Tables 1–4 list the accuracy, PPV, sensitivity, specificity, and *F1* scores of all models in terms of inhalation, exhalation, CAS, and DAS detection based on the HF_Lung_V1 dataset.

**Appendix A Table 1.**
**Accuracy, PPV, sensitivity, specificity, and *F1* scores of all models in inhalation detection.**

| Models | n of trainable parameters | Accuracy | | PPV | | Sensitivity | | Specificity | | *F1* score | |
|---|---|---|---|---|---|---|---|---|---|---|---|
| | | Segment Detection | Event Detection | Segment Detection | Event Detection | Segment Detection | Event Detection | Segment Detection | Event Detection | Segment Detection | Event Detection |
| LSTM | 300,609 | 0.890 | NA | 0.781 | 0.890 | 0.701 | 0.664 | 0.944 | NA | 0.739 | 0.761 |
| GRU | 227,265 | 0.899 | NA | 0.801 | 0.904 | 0.726 | 0.696 | 0.948 | NA | 0.762 | 0.789 |
| BiLSTM | 732,225 | 0.906 | NA | 0.814 | 0.885 | 0.750 | 0.772 | 0.951 | NA | 0.781 | 0.840 |
| BiGRU | 552,769 | 0.916 | NA | 0.836 | 0.898 | 0.773 | 0.800 | 0.956 | NA | 0.803 | 0.862 |
| CNN-LSTM | 3,448,513 | 0.903 | NA | 0.809 | 0.898 | 0.747 | 0.730 | 0.948 | NA | 0.776 | 0.811 |
| CNN-GRU | 2,605,249 | 0.905 | NA | 0.804 | 0.906 | 0.765 | 0.742 | 0.945 | NA | 0.784 | 0.820 |
| CNN-BiLSTM | 6,959,809 | 0.914 | NA | 0.822 | 0.902 | 0.791 | 0.803 | 0.950 | NA | 0.806 | 0.863 |
| CNN-BiGRU | 5,240,513 | 0.914 | NA | 0.829 | 0.898 | 0.785 | 0.812 | 0.952 | NA | 0.806 | 0.862 |
| SIMP BiLSTM | 235,073 | 0.906 | NA | 0.817 | 0.882 | 0.743 | 0.773 | 0.952 | NA | 0.778 | 0.841 |
| SIMP BiGRU | 178,113 | 0.915 | NA | 0.837 | 0.894 | 0.769 | 0.803 | 0.957 | NA | 0.801 | 0.861 |
| SIMP CNN-BiLSTM | 3,382,977 | 0.912 | NA | 0.828 | 0.895 | 0.774 | 0.799 | 0.953 | NA | 0.800 | 0.858 |
| SIMP CNN-BiGRU | 2,556,097 | 0.913 | NA | 0.830 | 0.889 | 0.774 | 0.810 | 0.953 | NA | 0.801 | 0.859 |

SIMP means the number of trainable parameters is adjusted.

## Appendix A Table 2.

## Accuracy, PPV, sensitivity, specificity, and *F1* scores of all models in exhalation detection.

| Models | n of trainable parameters | Accuracy | | PPV | | Sensitivity | | Specificity | | *F1* score | |
|---|---|---|---|---|---|---|---|---|---|---|---|
| | | Segment Detection | Event Detection | Segment Detection | Event Detection | Segment Detection | Event Detection | Segment Detection | Event Detection | Segment Detection | Event Detection |
| LSTM | 300,609 | 0.855 | NA | 0.716 | 0.561 | 0.406 | 0.456 | 0.962 | NA | 0.518 | 0.570 |
| GRU | 227,265 | 0.868 | NA | 0.715 | 0.687 | 0.514 | 0.554 | 0.951 | NA | 0.598 | 0.656 |
| BiLSTM | 732,225 | 0.866 | NA | 0.739 | 0.630 | 0.469 | 0.532 | 0.961 | NA | 0.573 | 0.639 |
| BiGRU | 552,769 | 0.882 | NA | 0.772 | 0.713 | 0.548 | 0.617 | 0.962 | NA | 0.641 | 0.709 |
| CNN-LSTM | 3,448,513 | 0.864 | NA | 0.732 | 0.628 | 0.476 | 0.512 | 0.957 | NA | 0.577 | 0.621 |
| CNN-GRU | 2,605,249 | 0.863 | NA | 0.731 | 0.629 | 0.470 | 0.516 | 0.958 | NA | 0.572 | 0.620 |
| CNN-BiLSTM | 6,959,809 | 0.867 | NA | 0.729 | 0.677 | 0.520 | 0.557 | 0.952 | NA | 0.604 | 0.656 |
| CNN-BiGRU | 5,240,513 | 0.874 | NA | 0.747 | 0.693 | 0.533 | 0.600 | 0.956 | NA | 0.622 | 0.685 |
| SIMP BiLSTM | 235,073 | 0.864 | NA | 0.736 | 0.612 | 0.450 | 0.520 | 0.962 | NA | 0.558 | 0.624 |
| SIMP BiGRU | 178,113 | 0.878 | NA | 0.741 | 0.716 | 0.559 | 0.603 | 0.954 | NA | 0.637 | 0.700 |
| SIMP CNN-BiLSTM | 3,382,977 | 0.869 | NA | 0.737 | 0.667 | 0.513 | 0.569 | 0.955 | NA | 0.604 | 0.662 |
| SIMP CNN-BiGRU | 2,556,097 | 0.873 | NA | 0.736 | 0.697 | 0.543 | 0.598 | 0.952 | NA | 0.624 | 0.684 |

SIMP means the number of trainable parameters is adjusted.

**Appendix A Table 3.**

**Accuracy, PPV, sensitivity, specificity, and *F1* scores of all models in CAS detection.**

| Models | n of trainable parameters | Accuracy | | PPV | | Sensitivity | | Specificity | | *F1* score | |
|---|---|---|---|---|---|---|---|---|---|---|---|
| | | Segment Detection | Event Detection | Segment Detection | Event Detection | Segment Detection | Event Detection | Segment Detection | Event Detection | Segment Detection | Event Detection |
| LSTM | 300,609 | 0.812 | NA | 0.554 | 0.120 | 0.087 | 0.095 | 0.983 | NA | 0.151 | 0.122 |
| GRU | 227,265 | 0.812 | NA | 0.529 | 0.217 | 0.160 | 0.153 | 0.966 | NA | 0.246 | 0.201 |
| BiLSTM | 732,225 | 0.815 | NA | 0.579 | 0.155 | 0.119 | 0.167 | 0.980 | NA | 0.198 | 0.191 |
| BiGRU | 552,769 | 0.818 | NA | 0.574 | 0.237 | 0.176 | 0.227 | 0.969 | NA | 0.269 | 0.256 |
| CNN-LSTM | 3,448,513 | 0.840 | NA | 0.676 | 0.475 | 0.341 | 0.329 | 0.960 | NA | 0.453 | 0.425 |
| CNN-GRU | 2,605,249 | 0.849 | NA | 0.689 | 0.556 | 0.411 | 0.402 | 0.955 | NA | 0.515 | 0.498 |
| CNN-BiLSTM | 6,959,809 | 0.844 | NA | 0.686 | 0.443 | 0.369 | 0.419 | 0.959 | NA | 0.479 | 0.464 |
| CNN-BiGRU | 5,240,513 | 0.851 | NA | 0.690 | 0.508 | 0.435 | 0.463 | 0.952 | NA | 0.533 | 0.516 |
| SIMP BiLSTM | 235,073 | 0.814 | NA | 0.560 | 0.152 | 0.121 | 0.148 | 0.977 | NA | 0.198 | 0.179 |
| SIMP BiGRU | 178,113 | 0.814 | NA | 0.546 | 0.202 | 0.162 | 0.178 | 0.968 | NA | 0.250 | 0.222 |
| SIMP CNN-BiLSTM | 3,382,977 | 0.848 | NA | 0.688 | 0.490 | 0.403 | 0.443 | 0.956 | NA | 0.508 | 0.502 |
| SIMP CNN-BiGRU | 2,556,097 | 0.851 | NA | 0.699 | 0.499 | 0.423 | 0.475 | 0.955 | NA | 0.526 | 0.515 |

SIMP means the number of trainable parameters is adjusted.

## Appendix A Table 4.

## Accuracy, PPV, sensitivity, specificity, and *F1* scores of all models in DAS detection.

| Models | n of trainable parameters | Accuracy | | PPV | | Sensitivity | | Specificity | | *F1* score | |
|---|---|---|---|---|---|---|---|---|---|---|---|
| | | Segment Detection | Event Detection | Segment Detection | Event Detection | Segment Detection | Event Detection | Segment Detection | Event Detection | Segment Detection | Event Detection |
| LSTM | 300,609 | 0.800 | NA | 0.716 | 0.699 | 0.556 | 0.485 | 0.905 | NA | 0.626 | 0.591 |
| GRU | 227,265 | 0.805 | NA | 0.697 | 0.746 | 0.624 | 0.514 | 0.883 | NA | 0.659 | 0.625 |
| BiLSTM | 732,225 | 0.821 | NA | 0.713 | 0.755 | 0.681 | 0.609 | 0.881 | NA | 0.696 | 0.700 |
| BiGRU | 552,769 | 0.827 | NA | 0.727 | 0.765 | 0.681 | 0.638 | 0.889 | NA | 0.703 | 0.714 |
| CNN-LSTM | 3,448,513 | 0.813 | NA | 0.706 | 0.734 | 0.672 | 0.526 | 0.876 | NA | 0.688 | 0.644 |
| CNN-GRU | 2,605,249 | 0.815 | NA | 0.725 | 0.709 | 0.640 | 0.539 | 0.893 | NA | 0.680 | 0.646 |
| CNN-BiLSTM | 6,959,809 | 0.830 | NA | 0.742 | 0.741 | 0.685 | 0.633 | 0.895 | NA | 0.712 | 0.708 |
| CNN-BiGRU | 5,240,513 | 0.826 | NA | 0.731 | 0.718 | 0.683 | 0.626 | 0.889 | NA | 0.706 | 0.700 |
| SIMP BiLSTM | 235,073 | 0.815 | NA | 0.702 | 0.764 | 0.675 | 0.582 | 0.876 | NA | 0.688 | 0.689 |
| SIMP BiGRU | 178,113 | 0.824 | NA | 0.718 | 0.779 | 0.689 | 0.626 | 0.883 | NA | 0.703 | 0.713 |
| SIMP CNN-BiLSTM | 3,382,977 | 0.828 | NA | 0.751 | 0.727 | 0.662 | 0.624 | 0.902 | NA | 0.702 | 0.702 |
| SIMP CNN-BiGRU | 2,556,097 | 0.825 | NA | 0.741 | 0.712 | 0.663 | 0.622 | 0.896 | NA | 0.699 | 0.695 |

SIMP means the number of trainable parameters is adjusted.

## Appendix B

### Supplementary Information regarding HF_Lung_V1

The open-access HF_Lung_V1 database, which comprises labels and lung sound recordings, can be found at https://gitlab.com/techsupportHF/HF_Lung_V1. The database is licensed under CC BY 4.0.

Detailed instruction about how to use the database can be found in the README.md in the repository.